\documentstyle[12pt,epsf]{article}

\def\hybrid{\topmargin 0pt      \oddsidemargin 0pt
	\headheight 0pt \headsep 0pt
	\textwidth 6.25in       
        \textheight 9.5in       
	\marginparwidth .875in
	\parskip 5pt plus 1pt   \jot = 1.5ex}

\catcode`\@=11
\def\marginnote#1{}

\newcount\hour
\newcount\minute
\newtoks\amorpm
\hour=\time\divide\hour by60
\minute=\time{\multiply\hour by60 \global\advance\minute by-\hour}
\edef\standardtime{{\ifnum\hour<12 \global\amorpm={am}%
	\else\global\amorpm={pm}\advance\hour by-12 \fi
	\ifnum\hour=0 \hour=12 \fi
	\number\hour:\ifnum\minute<10 0\fi\number\minute\the\amorpm}}
\edef\militarytime{\number\hour:\ifnum\minute<10 0\fi\number\minute}

\def\draftlabel#1{{\@bsphack\if@filesw {\let\thepage\relax
   \xdef\@gtempa{\write\@auxout{\string
      \newlabel{#1}{{\@currentlabel}{\thepage}}}}}\@gtempa
   \if@nobreak \ifvmode\nobreak\fi\fi\fi\@esphack}
	\gdef\@eqnlabel{#1}}
\def\@eqnlabel{}
\def\@vacuum{}
\def\draftmarginnote#1{\marginpar{\raggedright\scriptsize\tt#1}}

\def\draft{\oddsidemargin -.5truein
	\def\@oddfoot{\sl preliminary draft \hfil
	\rm\thepage\hfil\sl\today\quad\militarytime}
	\let\@evenfoot\@oddfoot \overfullrule 3pt
	\let\label=\draftlabel
	\let\marginnote=\draftmarginnote
   \def\@eqnnum{(\theequation)\rlap{\kern\marginparsep\tt\@eqnlabel}%
\global\let\@eqnlabel\@vacuum}  }

\def\numberbysection{\@addtoreset{equation}{section}
	\def\theequation{\thesection.\arabic{equation}}}

\catcode`@=12
\relax

\def\nn{\nonumber}
\def \Rs {\sf I\hskip-1.5pt R} 
\def\beq{\begin{equation}}
\def\eeq{\end{equation}}
\def\bea{\begin{eqnarray}}
\def\eea{\end{eqnarray}}

\relax
\numberbysection
\hybrid

\begin{document}
\begin{titlepage}
\begin{center}
{\large\bf On the universality of compact polymers}\\[.3in] 
        {\bf Jesper Lykke Jacobsen} \\ 
	{\it Laboratoire de Physique Statistique%
             \footnote{Laboratoire associ{\'e} aux universit{\'e}s
                       Paris 6, Paris 7 et au CNRS.},\\
             Ecole Normale Sup{\'e}rieure,\\
             24 rue Lhomond,
             F-75231 Paris CEDEX 05, FRANCE. \\}
\end{center}
\vskip .15in
\centerline{\bf ABSTRACT}
\begin{quotation}

{\small Fully packed loop models on the square and the honeycomb
lattice constitute new classes of critical behaviour, distinct from
those of the low-temperature O($n$) model. A simple symmetry argument
suggests that such compact phases are only possible when the
underlying lattice is bipartite. Motivated by the hope of identifying
further compact universality classes we therefore study the fully
packed loop model on the square-octagon lattice. Surprisingly, this
model is only critical for loop weights $n < 1.88$, and its scaling
limit coincides with the dense phase of the O($n$) model. For $n=2$ it
is exactly equivalent to the selfdual 9-state Potts model. These
analytical predictions are confirmed by numerical transfer matrix
results. Our conclusions extend to a large class of bipartite
decorated lattices.}

\end{quotation}
\end{titlepage}

\section{Introduction}

Compact polymers, the continuum limit of random walks that are
constrained to visit every site of some lattice ${\cal L}$, are
intriguing in so far as their critical exponents depend
explicitly on ${\cal L}$. Whilst first observed numerically
\cite{nien_fpl}, this curious lack of universality was firmly
established through the exact solution of the compact polymer problem
on the honeycomb \cite{batch94,jk_jpa} and, very recently, the square
lattice \cite{jj_npb,jj_prl}.

However, not every lattice can support a compact polymer phase.
To see this, consider more generally an O($n$)-type loop model defined on
${\cal L}$, in which each closed loop is weighed by $n$, and each
vertex {\em not} visited by a loop carries a factor of $t$. It is
well-known that for $|n| \le 2$ this model possesses a branch of
low-temperature ($t$ being the temperature) attractive critical fixed
points \cite{Nienhuis82,Blote89} with critical exponents that do not
depend on ${\cal L}$, even when ${\cal L}$ is not a regular
lattice but an arbitrary network \cite{DupSaleur87}. On the other
hand, whenever the model is invariant under $t \to -t$, as is the
case if ${\cal L}$ can only accommodate loops of {\em even} length,
this symmetry allows for a distinct zero-temperature branch of
repulsive fixed points \cite{nien_fpl}, with the $n \to 0$ limit
representing the compact polymer problem. That the critical behaviour
of this class of fully-packed loop (FPL) models depends on ${\cal L}$
is readily seen from the solutions of the honeycomb and the square case given
in Refs.~\cite{jk_jpa,jj_npb,jj_prl}. Namely, the continuum limit of
these models can be described by a conformal field theory (CFT) for a
fluctuating interface, where the fully-packing constraint 
forces the height variable to be a {\em vector}, with a number of
components that depends on the coordination number of the lattice at
hand.

This $t \to -t$ symmetry argument, originally put forward by Bl{\"o}te and
Nienhuis \cite{nien_fpl}, prompts us to conjecture its inverse: Whenever
${\cal L}$ allows for loops of {\em odd} length, so that the symmetry
is destroyed, the renormalisation group flow can be expected to take
us to non-zero $t$, eventually terminating in the 
dense, universal O($n$) phase. Support for this conjecture so far
comes from numerics in the case of
the triangular lattice \cite{batch96}, and recently for a class of
decorated lattices interpolating between the square and the triangular
lattices \cite{higuchi}.%
\footnote{Although belonging to the universality class of the square
lattice FPL model \cite{jj_npb,jj_prl} the FPL model on the
square-diagonal lattice does not constitute a very good
counterexample, since the fully-packing constraint actually prevents
the loops from occupying the diagonal edges. (Note that the proof
given in Ref.~\cite{higuchi} is also valid for $n \neq 0$.)}

Accepting for the moment the validity of this conjecture however
leaves us with an infinite set of bipartite lattices, each one being a
potential candidate for a novel universality class of compact polymers.
This perspective is especially appealing in the light of the constructive
point of view taken in Refs.~\cite{jk_jpa,jj_npb,jj_prl,jj_jsp}. In
these papers new CFTs were explicitly constructed,
based on purely geometrical considerations applied to the FPL
model in question. On the other hand, if the bipartite lattices
generate an entire family of distrinct CFTs, this gives rise to
important classification issues. In particular, the applicability of
compact polymer models to the protein folding problem
\cite{chan_dill,dill_rev} implies that one would like to understand on
which microscopic parametres (bending angles, coordination number,
steric constraints) the resulting conformational exponents do depend.

In this paper we examine FPL models on a class of
bipartite lattices, in which every vertex of a regular (square or
honeycomb) lattice has been decorated. A renormalisation group (RG)
argument, essentially amounting to a summation over the decoration,
reveals that the Liouville field theory construction \cite{jk_prl} 
should really be based on the undecorated lattice, but with bare
vertex weights that depend on the loop fugacity $n$. This leads to a
novel scenario in which, depending on $n$, the model may either
renormalise towards the dense phase of the O($n$) model or flow off to
a non-critical phase, even for $n<2$!

\begin{figure}
\begin{center}
\leavevmode
\epsfysize=200pt{\epsffile{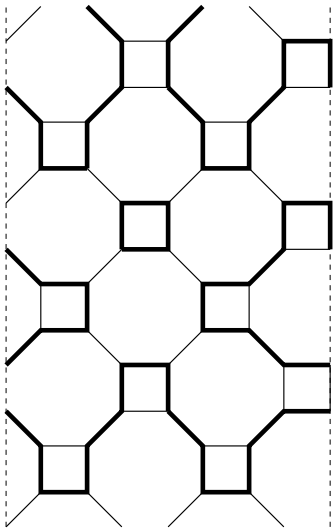}}
\end{center}
\protect\caption[2]{\label{fig:4-8}Fully packed loops on the
square-octagon lattice. In the corresponding transfer matrix, periodic
boundary conditions are imposed across a strip of width $L$ loop
segments (here $L=4$). The state space is that of all well-nested,
pairwise connections amongst the $L$ dangling ends in the upper row.}
\end{figure}

The case of the square-octagon lattice, shown in Fig.~\ref{fig:4-8},
is investigated in detail. This lattice can be thought
of as a square lattice in which each vertex has been decorated with a
tilted square. Our interest in the square-octagon lattice stems from the
fact that it is bipartite and has the same coordination number as the
honeycomb lattice, but enjoys the symmetry of the square lattice.
In particular it will enable us to assess whether the critical behaviour
of compact polymers on a lattice ${\cal L}$ depends only on its
coordination number%
\footnote{In the protein folding language this determines the
number of close contacts per monomer of the folded chain.},
only on the bond angles, or on a combination of both these parameters.
Our analysis suggests that the
corresponding FPL model belongs to the dense O($n$)
phase for $n < 1.88$, whilst for $n > 1.88$ a finite correlation
length is generated. For $n=2$ we show rigorously that the model is
equivalent to the (non-critical) 9-state Potts model. The analytical
results are confirmed by numerical transfer matrix calculations on
strips of width up to $L_{\rm max}=18$ loop segments.

Having introduced the models in Section~\ref{sec:models}, we present the
analytical results in Section~\ref{sec:analysis} and the numerics in
Section~\ref{sec:numerics}. Our results are discussed in
Section~\ref{sec:discussion}.

\section{The models}
\label{sec:models}

A fully packed loop (FPL) model on a lattice ${\cal L}$ is defined by the
partition function 
\beq
 Z_{\rm FPL} = \sum_{\cal G_{\rm FPL}} n^N,
 \label{Z-FPL}
\eeq
where the sum runs over all configurations ${\cal G_{\rm FPL}}$ of
closed loops drawn along the edges of ${\cal L}$ so that every vertex
is visited by a loop. Within a given configuration a weight $n$ is
given to each of its $N$ loops.

\begin{figure}
\begin{center}
\leavevmode
\epsfysize=200pt{\epsffile{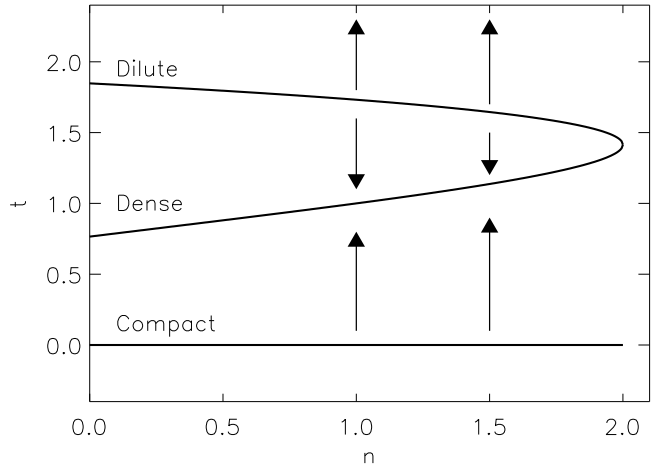}}
\end{center}
\protect\caption[2]{\label{fig:honeycomb}Phase diagram of the O($n$) model on
the honeycomb lattice.}
\end{figure}

An FPL model on ${\cal L}$ can be generalised to an O($n$) model by
lifting the fully packing constraint and further weighing each empty
vertex by a factor of $t$. Physically $t$ corresponds to a
temperature, the FPL model thus being the zero-temperature limit of
the O($n$) model. When ${\cal L}$ is the honeycomb lattice, the
resulting phase diagram is as shown on Fig.~\ref{fig:honeycomb}
\cite{nien_fpl}. For
$|n| \leq 2$ three branches, or phases, of critical behaviour
exist. Since ${\cal L}$ is bipartite, the resulting $t \to -t$
symmetry allows for a compact phase at 
$t=0$ \cite{nien_fpl,batch94,jk_jpa}, as discussed at length in the
Introduction. For $t>0$ Nienhuis has found the exact parametrisation
of a dense and a dilute phase, and determined the critical exponents
as a function of $n$ \cite{Nienhuis82}.

For our discussion of the square-octagon FPL model we shall need the
corresponding parametrisation for the O($n$) model on the {\em square}
lattice. The definition of the partition function is now slightly more
complicated, since each vertex can be visited by the loops in several
ways that are unrelated by rotational symmetry. An appropriate choice is
\beq
  Z_{{\rm O}(n)} = \sum_{\cal G} t^{N_t} u^{N_u} v^{N_v} w^{N_w} n^N,
  \label{Z-On}
\eeq
where $N_t$, $N_u$, $N_v$ and $N_w$ are the number of vertices visited
by respectively zero, one turning, one straight, and two mutually
avoiding loop segments. It is convenient to redefine the units of
temperature so that $t=1$.

Nienhuis \cite{Blote89,Batchelor89} has identified five branches of
critical behaviour for the model (\ref{Z-On}). The first four are
parametrised by 
\bea
  w_{\rm c} &=& \left \lbrace 2-\left[1-2 \sin \left(
                \frac \theta 2 \right) \right]
                \left[ 1+2 \sin \left(\frac \theta 2 \right) \right]^2
                \right \rbrace^{-1}, \nn \\
  u_{\rm c} &=& 4 w_{\rm c} \sin \left(\frac \theta 2 \right)
                \cos \left(\frac \pi 4 - \frac \theta 4 \right), \nn \\
  v_{\rm c} &=& w_{\rm c} \left[ 1+2 \sin \left( \frac \theta 2
                \right) \right], \nn \\ 
  n &=& -2 \cos (2\theta),
  \label{weights}
\eea
where $\theta \in [(2-b)\pi/2,(3-b)\pi/2]$ corresponds to branch
$b=1,2,3,4$. It has recently been noticed that the edges {\em not}
covered by the original (`black') loops form a second species of
closed (`grey') loops, each one occuring with unit weight
\cite{jj_jsp}. Lifting the fully-packing constraint implies that the
two loop flavours decouple, and each of them can independently reside
in either of the two critical phases (dense or dilute) discussed
above. The black (resp.~grey) loops are dense on branches 2 and 4
(resp.~1 and 2), and dilute on branches 1 and 3 (resp.~3 and 4).
On branches 1 and 2 the grey loops contribute neighter to the central
charge, nor to the geometrical (string) scaling dimensions, and in the
scaling limit these two branches are thus completely analogous to the
dilute and the dense branches of the O($n$) model on the honeycomb
lattice \cite{jj_jsp}.

The last critical branch, known as branch 0, has weights
\beq
  u_{\rm c} = w_{\rm c} = \frac12, \ \ \ \ v_{\rm c} = 0, \ \ \ \
  -3 \leq n \leq 1,
  \label{branch0}
\eeq
and can be exactly mapped onto the dense phase of the O($n+1$) model
\cite{Blote89}, or equivalently to the selfdual $(n+1)^2$-state Potts
model \cite{Nienhuis82}.

\section{Renormalisation group analysis and an exact mapping}
\label{sec:analysis}

At first sight it would seem that the continuum limit of the FPL model
(\ref{Z-FPL}) on the square-octagon lattice should be described by a
Liouville field theory for a two-dimensional height field, since the
lattice has the same coordination number as the honeycomb lattice
\cite{jk_jpa}. However, we shall presently see that only {\em one}
height component survives when applying the appropriate coarse
graining procedure to the two-dimensional microscopic heights defined
on the lattice plaquettes.

\begin{figure}
\begin{center}
\leavevmode
\epsfysize=200pt{\epsffile{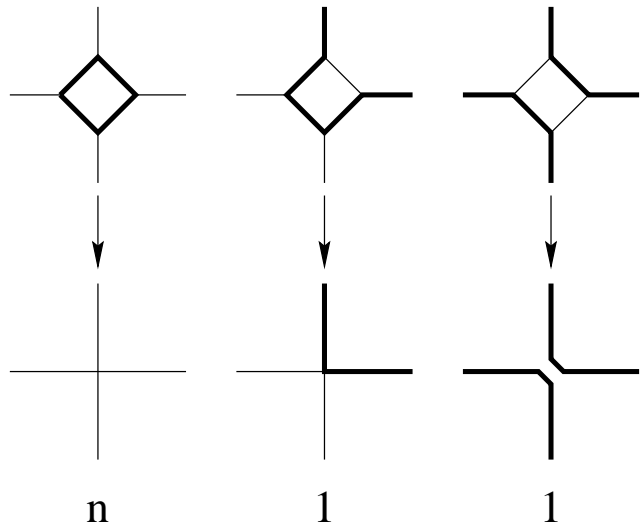}}
\end{center}
\protect\caption[2]{\label{fig:real-RG}First step in a real-space
renormalisation of the square-octagon lattice FPL model. The
renormalised vertices get weighted as shown.}
\end{figure}

Consider performing the first step of a real-space renormalisation group
(RG) transformation of Eq.~(\ref{Z-FPL}), by summing over the degrees of
freedom residing at the decorating squares. In this way the decorated
vertices transform into weighted undecorated vertices, as shown on
Fig.~\ref{fig:real-RG}. The renormalised model is then simply the
O($n$) model on the square lattice (\ref{Z-On}), but with some
particular `bare' values of the vertex weights. Defining again the
empty vertex to have unit weight, these bare weights read
\beq
  u = \frac{1}{n}, \ \ \ \
  v = 0, \ \ \ \
  w = \frac{1}{n}.
  \label{bare}
\eeq

Following the standard procedure \cite{jk_npb} microscopic heights can
be defined on the lattice plaquettes by orienting the loops and
assigning a vector, ${\bf A}$, ${\bf B}$ or ${\bf C}$, to each of the
three possible bond states: ${\bf A}$ (${\bf B}$) if the bond is
covered by a loop directed towards (away from) a site of the even
sublattice, and ${\bf C}$ if the bond is empty. When encircling an
even (odd) site in the (counter)clockwise direction the microscopic
height increases by the corresponding vector whenever a bond is
crossed. As was first pointed out in Ref.~\cite{jk_jpa} the
fully-packing constraint leads to the condition
${\bf A} + {\bf B} + {\bf C}={\bf 0}$, whence the height must a priori
be two-dimensional. However, the RG transformation that we have just
applied lifts the fully-packing constraint, due to the appearance of the
bottom left vertex of Fig.~\ref{fig:real-RG}. Defining now the
sublattices with respect to the renormalised (square) lattice we have
the additional constraint $4 {\bf C} = {\bf 0}$, whence the coarse
grained height field should really be one-dimensional%
\footnote{See Ref.~\cite{jj_jsp} for similar examples of such a
reduction of the dimensionality of the height field.},
and O($n$)-like behaviour is to be expected. Also note that it clearly
suffices to define the microscopic 
heights on the octagonal plaquettes in order to obtain a continuous
height field defined everywhere in $\Rs^2$ by the usual coarse graining
procedure \cite{jk_npb}.

The reason that the renormalised FPL model is still interesting is
that the bare vertex weights (\ref{bare}) are now some fixed functions
of the loop fugacity $n$, rather than arbitrary parameters that can be
tuned to their critical values. This constitutes an interesting 
situation which has not been encountered before. We shall soon see
that it implies that the FPL model (\ref{Z-FPL}), unlike any other
loop model studied this far, is only critical within a part of the
interval $|n| \leq 2$.

\begin{figure}
\begin{center}
\leavevmode
\epsfysize=200pt{\epsffile{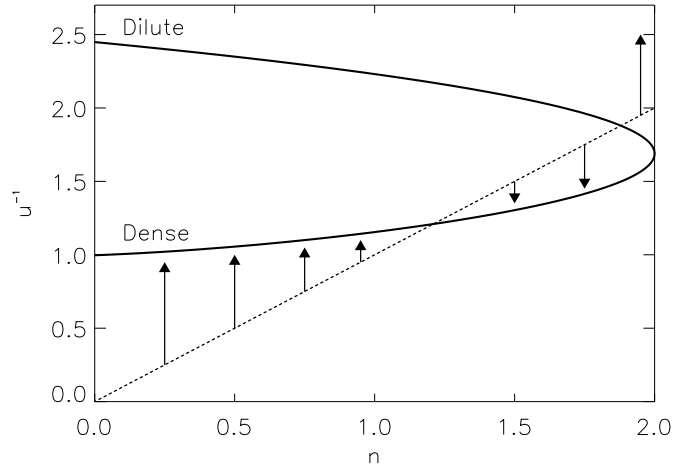}}
\end{center}
\protect\caption[2]{\label{fig:flow}RG flow in the square-octagon FPL
model. After tracing over the decoration, the bare value of $1/u$ is
given by the dashed line. For $n<1.88$ the flow is directed towards
the attractive branch of dense fixed points, whilst for $1.88 < n < 2$
the system renormalises towards the high-temperature disordered phase.}
\end{figure}

In Fig.~\ref{fig:flow} we show $1/u_{\rm c}$, the weight of the empty vertex
relative to that of a turning loop segment, as a function of $n$ for
the critical branches 1 (dilute phase) and 2 (dense phase) of the
O($n$) model on the square lattice; cfr.~Eq.~(\ref{weights}). In analogy
with the honeycomb case the dense and dilute branches again consist of
respectively attractive and repulsive fixed points. With the bare
value $1/u$ given by Eq.~(\ref{bare}) the subsequent RG flow must
therefore be as schematically indicated on the figure.
For $n \simeq 1.88$ there is an intersection between the bare value
and that of the dilute branch, and for $n > 1.88$ we can therefore
expect the flow to be 
directed towards the high-temperature disordered phase of the O($n$)
model. In other words, a finite correlation length (roughly the size
of the largest loop in a typical configuration) is generated and
the model is no longer critical.

Of course we should be a little more careful, since $u$ is not the
only parameter in the model. Whenever the bare weights (\ref{bare}) do
not intersect one of the five branches of fixed points, $v$ and $w$
will flow as well. In particular, $v$ will in general flow towards
non-zero values, since the turning loop segments always occur with finite
weight, and these are clearly capable of generating straight loop
segments on larger length scales. The essential point is that for
$1.88 < n < 2$ empty vertices will begin to proliferate, and there is
no physical mechanism for halting the flow towards the disordered
phase.%
\footnote{The flow cannot be towards branch 0 since this is a
repulsive fixed point.}

The point $n=2$ merits special attention. Here the bare weights are
\beq
  u = w = \frac12, \ \ \ \ v = 0,
\eeq
which coincides with the fixed point values on branch 0;
see Eq.~(\ref{branch0}). Invoking Nienhuis' mapping \cite{Blote89} the
$n=2$ FPL model is therefore exactly equivalent to the selfdual
9-state Potts model, which is of course again non-critical \cite{Baxter73}.

\section{Transfer matrix results}
\label{sec:numerics}

In order to confirm the analytical predictions given in Section
\ref{sec:analysis} we have numerically calculated effective values of the
central charge $c$ and the thermal scaling dimension $x_t$ on strips
of width $L=4,6,\ldots,18$ loop segments. To this end we adapted the
connectivity basis transfer matrices described in
Refs.~\cite{Blote89,jj_npb} to the square-octagon lattice. The working
principle of these transfer matrices is illustrated in
Fig.~\ref{fig:4-8}: To determine the number of loop closures induced
by the addition of a new row of vertices it suffices to know the
pairwise connections amongst the $L$ dangling ends of the top row.
For $L$ even, the number of such connections is \cite{Blote89}
\beq
 a_L = \sum_{i=0}^{L/2} {L \choose 2i} c_{L/2-i},
\eeq
where $c_m = \frac{(2m)!}{m!(m+1)!}$ are the Catalan numbers. Thus,
the transfer matrix for a strip of width $L$ has dimensions
$a_L \times a_L$, and a sparse matrix decomposition can be made by
adding one site of the lattice at a time, rather than an entire row.
The size of the largest matrix employed is given by $a_{18} = 6,536,382$.

\begin{table}
\begin{center}
\begin{tabular}{|r|rrrrrr|r|} \hline
$n$ & $c(4,8)$ & $c(6,10)$ & $c(8,12)$ & $c(10,14)$ & $c(12,16)$ &
$c(14,18)$ & O($n$) \\ \hline
0.0 & -1.9784 & -1.9862 & -1.9880 & -1.9924 & -1.9963 & -1.9980 & -2.0000 \\
0.5 & -0.8898 & -0.8975 & -0.8729 & -0.8488 & -0.8338 & -0.8259 & -0.8197 \\
1.0 &  0.0706 & -0.0856 & -0.0965 & -0.0701 & -0.0434 & -0.0246 &  0.0000 \\
1.5 &  0.9390 &  0.6602 &  0.5504 &  0.5144 &  0.5117 &  0.5227 & 0.5876 \\
2.0 &  1.6484 &  1.4844 &  1.4495 &  1.4362 &  1.4225 &  1.4068 & 1.0000 \\
\hline
\end{tabular}
\end{center}
\caption{Three-point estimates for the central charge, compared with
exact results for the dense phase of the O($n$) model.}
\label{tab:c}
\end{table}

The effective central charge $c(L,L+4)$ has been estimated by
three-point fits of the form \cite{bcn,affleck,jj_npb}
\begin{equation}
  f_0(L) = f_0(\infty) - \frac{\pi c}{6 L^2} + \frac{A}{L^4} + \cdots
  \label{fc2}
\end{equation}
applied to the free energy per site $f_0(L')$ with $L'=L,L+2,L+4$.
Similarly, effective values $x_t(L,L+2)$ of the thermal scaling
dimension were found from two-point fits of the form \cite{Cardy83,jj_npb}
\begin{equation}
 \label{fx2}
 f_1(L) - f_0(L) = \frac{2\pi x_t}{L^2} + \frac{B}{L^4} + \cdots,
\end{equation}
where $f_0(L)$ and $f_1(L)$ are related to the ground state and the
first excited state of the transfer matrix spectra in the usual way.

The numerical results are given in Tables~\ref{tab:c} and
\ref{tab:xt}. For $n \leq 1.5$ we see the expected convergence towards
the exact values of the O($n$) model in the dense phase, which read
\cite{nien_rev}
\beq
  c = 1 - \frac{6 e^2}{1-e}, \ \ \ \
  x_t = \frac{2e+1}{2(1-e)}
\eeq
with $e \equiv \frac{1}{\pi}\arccos(n/2)$. For $n=1.5$ the convergence
is rather slow, especially in the case of $c$, reflecting a large
crossover length.

As predicted by theory, the FPL model is no longer critical at
$n=2$. This is particulary visible from the monotonic decrease of the
$x_t$ estimates, which are well below the exact O($n$) value $x_t=1/2$.
For a system with a finite correlation length, $\xi < \infty$, the
effective values for $c$ should eventually tend to zero. The fact that
we observe rather large effective values is in agreement with
Ref.~\cite{Blote89}, and rather predictable since $\xi$ is much
greater than the largest strip width used in the simulations
\cite{Buffenoir}.
For comparison we performed similar computations for
the 9-state Potts model in its loop representation \cite{3-Potts},
finding again effective values of $c$ in the range 1.3--1.4.

\begin{table}
\begin{center}
\begin{tabular}{|r|rrrrrrr|r|} \hline
$n$ &$x_t(4,6)$ & $x_t(6,8)$ & $x_t(8,10)$ & $x_t(10,12)$ &
$x_t(12,14)$ & $x_t(14,16)$ & $x_t(16,18)$ & O($n$) \\ \hline
0.0 & 1.2942 & 1.9142 & 2.5336 & 2.5970 & 1.9830 & 1.9849 & 1.9890 & 2.0000 \\
0.5 & 1.3792 & 1.8833 & 2.4173 & 1.5522 & 1.5201 & 1.5557 & 1.5685 & 1.5843 \\
1.0 & 1.2480 & 1.2843 & 1.2895 & 1.2856 & 1.2799 & 1.2745 & 1.2701 & 1.2500 \\
1.5 & 0.8593 & 0.7842 & 0.8088 & 0.8392 & 0.8664 & 0.8876 & 0.9030 & 0.9482 \\
2.0 & 0.5488 & 0.5708 & 0.5629 & 0.4188 & 0.3864 & 0.3595 & 0.3369 & 0.5000 \\
\hline
\end{tabular}
\end{center}
\caption{Two-point estimates for the thermal scaling dimension,
juxtaposed with exact values for the dense O($n$) model.}
\label{tab:xt}
\end{table}

\section{Discussion}
\label{sec:discussion}

Having seen that two of the simplest two-dimensional lattices (square
and honeycomb) give rise to distinct compact universality classes, it
would be tempting to conjecture that an FPL model defined on any new
lattice leads to different critical exponents and has a new CFT
describing its continuum limit. In the present paper we have
demonstrated that this is far from being the case. Even 
within the very restricted class of bipartite lattices, fulfilling the
$t \to -t$ symmetry requirement, any lattice that can be viewed as a
decorated square or honeycomb lattice is likely to flow away from
the compact phase by virtue of an RG transformation analogous to the
one presented in Section \ref{sec:analysis}.

Despite the curious lattice dependence of the compact phases, it
thus appears that the number of distinct universality classes is very
restricted. We recall that the continuum limit of all loop models
solved to this date can be constructed by perturbing a
${\rm SU}(N)_{k=1}$ Wess-Zumino-Witten model by exactly marginal
operators and introducing an appropriate background charge
\cite{jk_npb}. It would  be most interesting to pursue the physical
reason why only the cases $N=2$ (the O($n$) \cite{jj_jsp},
Potts \cite{jk_conf} and six-vertex \cite{jk_npb} models),
$N=3$ (the FPL model on the honeycomb lattice \cite{jk_jpa}),
and $N=4$ (the two-flavoured FPL model on the square lattice
\cite{jj_npb,jj_prl}) seem to occur in practice.

The square-octagon lattice FPL model studied here turned out
to be interesting in several respects. First, it provides us with the
first example of an non-oriented \cite{Manhat}, bipartite
\cite{batch96,higuchi} lattice for which the scaling properties of
compact and dense polymers are identical. In particular, the exact
value of the conformational exponent $\gamma$ is $19/16$
\cite{nien_rev,DupSaleur87}, indicating a rather strong entropic
repulsion between the chain ends. Second, the square-octagon model
presents a novel scenario in which the same fully packed loop model
may renormalise towards different conformal field theories, or even
flow off to a non-critical regime, depending on the value of the loop
fugacity $|n| \leq 2$. In particular one might be able to `design' a
decorated lattice with bare vertex weights that simultaneously
intersect those of the dilute O($n$) phase for some value of $n$.
This could be a starting point
for gaining a microscopic, geometrical understanding of the Coulomb gas
charge asymmetry \cite{nien_rev} which was shown in Ref.~\cite{jj_jsp}
to distinguish between the dense and dilute phase of the O($n$) model.
Finally, our model proves that the scaling properties of compact
polymers do not depend exclusively on either bond angles or
coordination number, but rather on a combination of these two
parameters.

\noindent{\large\bf Acknowledgments}

The author is greatly indebted to J.~Kondev for many valuable comments
and suggestions, and would like to thank Saint Maclou for inspiration
during the initial stages of this project.

\end{document}